**Częstochowski**

**Kalendarz Astronomiczny**

**2013**

**Rok IX**

**Redakcja**

**Bogdan Wszołek**
**Agnieszka Kuźmicz**

**Wersja elektroniczna kalendarza jest dostępna na stronach**

**www.astronomianova.org**
**www.ptma.ajd.czest.pl**

# Period variations the intermediate polars EX Hya, FO Aqr and RXS J180340.0+401214

Vitalii V. Breus[1], Ivan L. Andronov[1], Tamas Hegedus[2], Pavol A. Dubovsky[3], Igor Kudzej[3], Karol Petrik[4], Stanisław Zoła[5,6]

[1] Department "High and Applied Mathematics", Odessa National Maritime University, Odessa, Ukraine
[2] Baja Astronomical Observatory, Baja, Hungary
[3] Vihorlat Astronomical Observatory, Humenne, Slovakia
[4] Astronomical Observatory, Hlohovec, Slovakia
[5] Astronomical Observatory, Jagiellonian Observatory, Cracow, Poland
[6] Mt. Suhora Observatory, Pedagogical University, Cracow, Poland

We present results of CCD photometric study of the intermediate polars FO Aqr, EX Hya and RXS J180340.0+401214.

EX Hya was observed using the telescopes of the Tzec Maun observatory in 2010 and 2011. Also we used observations obtained from ASAS, AAVSO and WASP data archives. Periodogram analysis was carried out. We analyzed variations of the spin period of the white dwarf in this system using published earlier and our spin maxima timings (total 451 moments since 1962).

Using the most recent spin variability characteristics published by [1] $P_0=0.046546484$, $T_0=2437699.8920$ we obtained new ephemeris

$$T_{max}=2437699.89079(59)+0.0465464808(69)\cdot E - 6.3(2)\cdot 10^{-13} E^2$$

which corresponds to the characteristic time of acceleration of rotation (spin-up) of $4.67(14)\cdot 10^6$ years.

The orbital variability of EX Hya is more complicated. The O-C diagram shows spin-up from the beginning of observations until 1978, then it changed with spin-down. Since 1986 till 2007 it showed spin-up again. This may indicate the existence of a third component of the system.

FO Aqr was observed using 50-cm telescope at the Baja Astronomical Observatory, Hungary. Additionally, we analyzed observations obtained in the Vihorlat Astronomical Observatory, Slovakia. The spin period during our observations was $0.^d014521(3)$ with an initial epoch for the maximum brightness of 2455068.72430(36). The best fit value of the orbital period of the system is $0.^d2120801$. This value corresponds to our light curve better than the published earlier value of $0.^d2020596$ [2], although the difference is significant.

The spin period variations of FO Aqr are complicated. From 1981 to 1987, the white dwarf showed a spin-down, then it changed to a spin-up. Due to a gap in the observations for almost 6 years there was a cycle miscounting, so we have 2 branches on the O-C diagram, and there is no published information which could help in filling this gap with points to restore the cycle numbering. This shows a very high importance of regular studies of such short period objects.

The newly discovered intermediate polar RXS J180340.0+401214 (RXJ1803) [3] was started to be observed using the Zeiss-Cassegrain 600 telescope of the Hlohovec Observatory and Planetarium, and then in the Vihorlat, Baja and Cracow Observatories.

The orbital variability is almost absent in our observations, no eclipses were found, suggesting a low orbit inclination $i<70^o$,

The photometric wave is originated due to a spin rotation of the white dwarf, during which the viewing conditions of the accretion columns are continuously changing. Therefore, the variability seems to be due to the geometric conditions (changing of the angle between stream and beam of view in the rotation), rather than for the physical ones (instability of the accretion column – that really is present, but not periodic). One hump shape at the phase light curve argues for a high inclination of the magnetic axis in this system, so we see mainly an upper accretion column.

The O-C analysis shows the necessity of improvement of the value of the spin period of 1520.4509±0.0022 seconds (25.34 minutes) obtained from our first observations consistent with that published earlier [3]. However, due to an error in the timing of Teichgraeber et al. [4] (they published epoch of minimum instead of maximum), previous attempt to fit all timings (Andronov et al. [5]) were unsuccessful. In this work, we present a new ephemeris for the spin maxima:

$$T_{max}= 2454604.04449(14)+0.017596986(3) \cdot E$$

The quadratic term $Q=(9 \pm 5)10^{-14}$ formally corresponds to characteristic time scale of period variations of $\tau=P/(dP/dt)=(4.6\pm2.5)$ Myr. However, the parameter is equal to 1.9 of its error estimate and thus is not statistically significant. Contrary to other intermediate polars, no period variations were detected in RXJ1803.

The color index shows a statistically significant dependence on the spin phase, indicating a variable distribution of energy in the spectrum and necessity of multicolor observations rather than mono-filter or unfiltered ones.